\documentclass[smallabstract,smallcaptions]{dccpaper}

\usepackage{epsfig}
\usepackage{citesort}
\usepackage{amsmath}
\usepackage{amssymb}
\usepackage{color}
\usepackage{url}
\usepackage{booktabs}
\usepackage{multirow}
\usepackage{caption}
\usepackage{subcaption}
\usepackage{placeins}
\usepackage{textcomp}

\usepackage{footmisc}

\newlength{\figurewidth}
\newlength{\smallfigurewidth}

\begin{document}

\title
{\large
\textbf{Rethinking Bjøntegaard Delta for Compression Efficiency Evaluation: Are We Calculating It Precisely and Reliably?}
}

\author{%
Xinyu Hang$^{\ast}$, Shenpeng Song$^{\ast}$, Zhimeng Huang$^{\ast}$, \\ Chuanmin Jia$^{\sharp}$, Siwei Ma$^{\ast\dag}$ and Wen Gao$^{\ast\dag}$\\[0.5em]
{\small\begin{minipage}{\linewidth}\begin{center}
\begin{tabular}{c}
$^{\ast}$School of Computer Science, Peking University, Beijing, China \\
$^{\sharp}$Wangxuan Institute of Computer Technology, Peking University, Beijing, China\\
$^{\dag}$Peng Cheng Laboratory, Shenzhen, China \\
\end{tabular}
\end{center}\end{minipage}}
}

\maketitle
\thispagestyle{empty}
\let\thefootnote\relax\footnotetext{Correspondence to: Zhimeng Huang (zmhuang@pku.edu.cn).}

\begin{abstract}

For decades, the Bjøntegaard Delta (BD) has been the metric for evaluating codec Rate-Distortion (R-D) performance. Yet, in most studies, BD is determined using just 4-5 R-D data points, could this be sufficient? As codecs and quality metrics advance, does the conventional BD estimation still hold up? Crucially, are the performance improvements of new codecs and tools genuine, or merely artifacts of estimation flaws? This paper addresses these concerns by reevaluating BD estimation. We present a novel approach employing a parameterized deep neural network to model R-D curves with high precision across various metrics, accompanied by a comprehensive R-D dataset. This approach both assesses the reliability of BD calculations and serves as a precise BD estimator. Our findings advocate for the adoption of rigorous R-D sampling and reliability metrics in future compression research to ensure the validity and reliability of results.

\end{abstract}

\Section{Introduction}

Over the past three decades, significant advancements have been made in multimedia coding, cementing its role as a cornerstone of modern technology. To quantitatively measure coding performance in a scalar metric, the BD metric~\cite{bjontegaard2001calculation}, introduced in 2001, has since become the gold standard for comparing the relative rate-distortion (R-D) performance of different codecs.

\begin{figure}
    \centering
    \begin{subfigure}[b]{0.5\textwidth}
        \includegraphics[width=\linewidth]{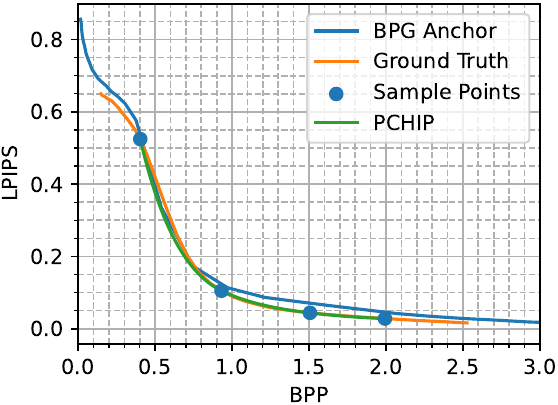}
        \caption{}
        \label{fig:sample-lpips}
    \end{subfigure}
    ~~~~~
    \begin{subfigure}[b]{0.4\textwidth}
        \includegraphics[width=\linewidth]{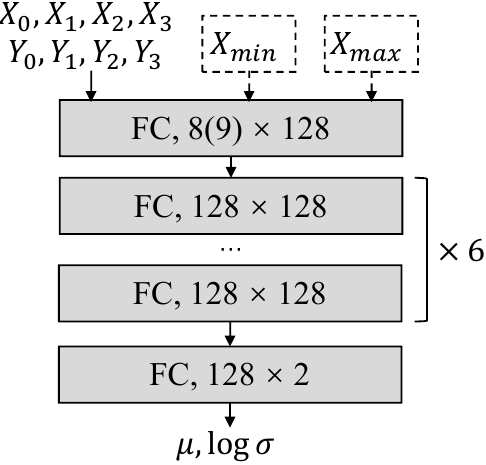}
        \caption{}
        \label{fig:mlp}
    \end{subfigure}
    
    \caption{(a) \textbf{Example of BD Estimation.} In this example, the true BD-BR (yellow line) is -2.078\%, while the generalized PCHIP method (green line) yields -9.77\%, resulting in a significant bias of 7.70\%. (b) \textbf{Integral Prediction Network.} ``FC, $x \times y$'' denotes a Fully-Connected layer with $x$ input elements and $y$ output elements.}
\end{figure}

BD is derived from approximating R-D curves using sample points, as acquiring accurate R-D curves is typically expensive. Initially, BD estimation was designed for traditional codecs using Peak Signal-to-Noise Ratio (PSNR) for distortion measurement, where a bilinear or cubic function provided a reasonable fit to the R-D curve, requiring only 4-5 sample points. Later, more sophisticated interpolation methods such as Cubic Spline Interpolation (CSI), Piecewise Cubic Hermite Interpolating Polynomial (PCHIP), and Akima spline have been utilized to achieve high-fidelity R-D curve fitting~\cite{herglotz2022beyond}. However, we find that the evolution of codec technology, especially the rise of learning-based codecs and complex quality metrics, has made BD estimation complicated, making existing methods increasingly inadequate. An illustrative example is shown in Fig.~\ref{fig:sample-lpips}. In this instance, the curve derived from the PCHIP interpolation starkly deviates from the actual R-D curve, which is determined through dense sampling across the R-D Laplacian parameter $\lambda$. This significant discrepancy has led to a BD-bitrate (BD-BR) estimation bias exceeding 7\%.

The implications of such phenomena are indeed serious. Typically, enhancements in codec performance result from incremental advancements introduced by various coding tools. However, potential bias in BD estimation may obscure the actual performance gains achieved by these tools. In such cases, it becomes difficult to determine whether the coding technology has genuinely led to substantial improvements.

In light of these challenges and limitations, we introduce a robust method for high-precision BD estimation across diverse compression scenarios, enhanced by a reliability assessment to determine the probability distribution of BD values from R-D sample points. Our method's validity is confirmed through extensive testing on a dataset we constructed. The key contributions can be summarized as follows:
\begin{itemize}
    \item We have established a large-scale, high-precision R-D dataset to verify the accuracy of existing BD estimation algorithms.
    \item We introduce a novel BD computation technique that not only offers a confidence measure for the predicted BD value but also functions as a high-accuracy BD predictor itself.
    \item We demonstrate the superiority of the framework in our evaluation dataset, showing reasonable confidence predictions as well as BD estimates with higher accuracy.
\end{itemize}

\Section{Physical and Mathematical Concept of BD}

The calculation of BD is based on multiple sample points selected from the rate-distortion (R-D) curves of two codecs: the anchor codec, represented by its rate \( R_A \) and distortion \( D_A \), and the target codec, represented by its rate \( R_B \) and distortion \( D_B \). For a reliable estimation of BD, at least four sample points are required on the R-D curves. The calculation proceeds as follows:

\textbf{Step 1:} Convert the bitrates to a logarithmic scale to equally weigh different rate intervals, preventing bias toward higher bitrates:
\begin{align}
    r_A = \log R_A, \quad r_B = \log R_B.
\end{align}

\textbf{Step 2:} Apply interpolation to fit the R-D curves of both codecs using the sample points \( (r, D) = \{ (r_i, D_i) \mid r_i \in \mathbb{R}, D_i \in \mathbb{R}, i = 1, 2, \ldots, N \} \). This results in \( r \) as a function of \( D \) and \( D \) as a function of \( r \), denoted as \( \hat{r}_A(D) \), \( \hat{D}_A(r) \), \( \hat{r}_B(D) \), and \( \hat{D}_B(r) \).

\textbf{Step 3:} Define the measurement intervals for BD-BR and BD-quality. Since the R-D curve interpolations often do not identically overlap, the valid intervals are given by the intersection of the two R-D curves' axes projections. For BD-BR, the interval is:
\begin{align}
    [D_{\text{min}}, D_{\text{max}}] = \left[\max(\min (D_A), \min (D_B)), \min(\max (D_A), \max (D_B))\right].
\end{align}
For BD-quality, the interval is:
\begin{align}
    [r_{\text{min}}, r_{\text{max}}] = \left[\max(\min (r_A), \min (r_B)), \min(\max (r_A), \max (r_B))\right].
\end{align}

\textbf{Step 4:} Calculate BD-BR (also called BD-rate), \( \Delta R \), as:
\begin{align}
    \Delta R = e^{\Delta r} - 1, \quad \Delta r = \frac{1}{D_{\text{max}} - D_{\text{min}}} \int_{D_{\text{min}}}^{D_{\text{max}}} \left( \hat{r}_B(x) - \hat{r}_A(x) \right) \, \text{d}x,
\end{align}
which measures the average bitrate reduction of codec B compared to codec A for the same distortion.
The BD-quality, \( \Delta D \), is similarly calculated as:
\begin{align}
    \Delta D &= \frac{1}{r_{\text{max}} - r_{\text{min}}} \int_{r_{\text{min}}}^{r_{\text{max}}} \left( \hat{D}_B(x) - \hat{D}_A(x) \right) \, \text{d}x,
\end{align}
describing the average change in distortion between codecs for a given bitrate, also on a logarithmic scale. Typically, the term ``BD-quality'' is replaced with the corresponding quality assessment indicator, such as ``BD-PSNR'', to denote the calculation of the change in the respective quality metric under the same conditions.

A recent work~\cite{bdbible} provides a comprehensive comparison of existing tools for calculating BD, with the primary distinction being the R-D interpolation method used in the second step. The most commonly used algorithms include cubic function fitting, CSI, PCHIP, and Akima spline. However, these methods cannot completely describe the difference between the difference between the fitted curve and the actual one, particularly evident when the number of sample points is insufficient. Moreover, the lack of confidence prevents a quantitative assessment of the estimation bias in BD. To address these issues, we pioneered the Bjøntegaard Delta Confidence Interval (BDCI) for quantitatively assessing BD measurement errors.

\Section{Our Approach: New Insights Named Bj{\o}ntegaard Delta
Confidence Interval}


A key insight is that calculating BD does not necessitate explicitly fitting an R-D curve. The sole purpose of the curve is to facilitate the computation of a definite integral over a specified interval along the relevant axis. Therefore, our method eliminates the need for curve fitting and directly estimates the integral using the sample points.

\begin{figure}[t]
    \centering
    \includegraphics[width=0.75\linewidth]{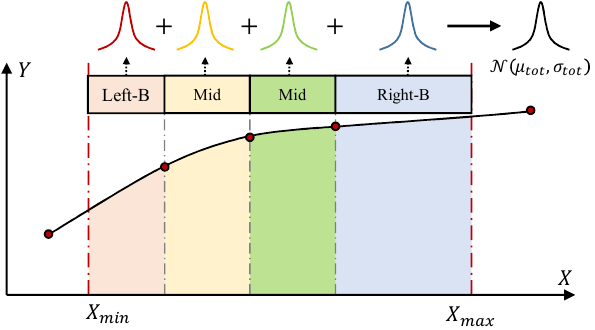}
    \caption{\textbf{Integral Estimation Example.} The integration interval is segmented into distinct segments based on the \( X \)-values of the sample points. For each segment, integrals are estimated, and these predicted values are subsequently aggregated.}
    \label{fig:estimate_overview}
\end{figure}

\begin{figure}[h]
    \centering
    \includegraphics[width=0.9\linewidth]{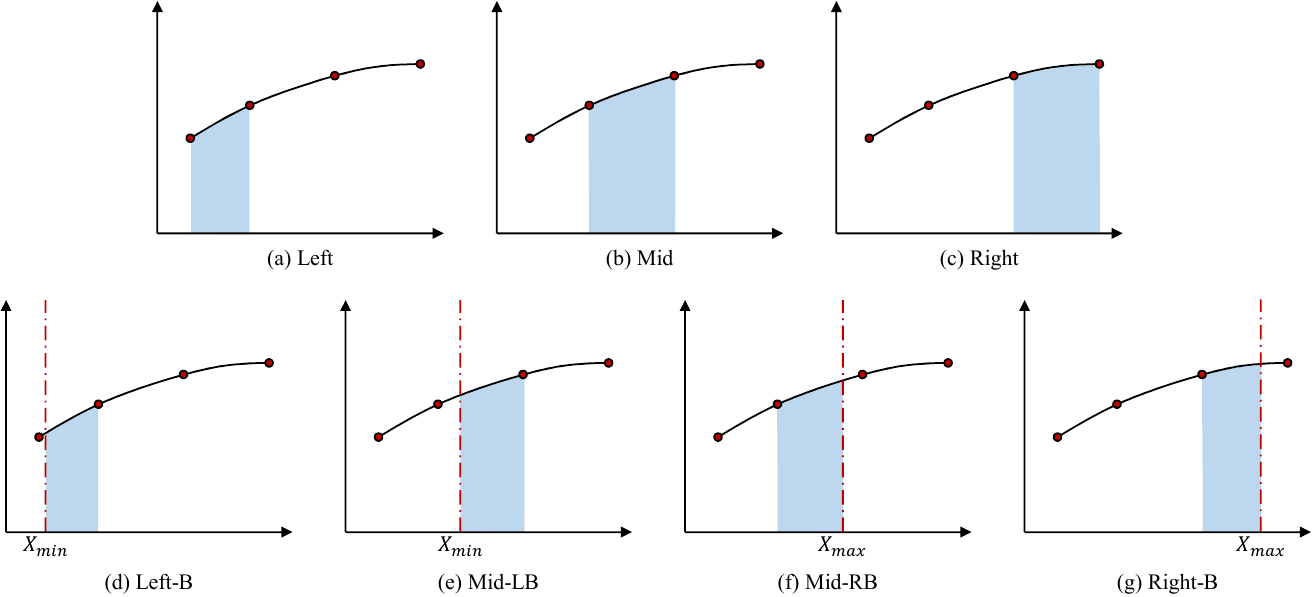}
    \caption{\textbf{Categories of Segments.} In total, there are 7 distinct categories of segments, each corresponding to a separate prediction model. The blue areas represent the target integration regions. The red dots and dashed lines indicate the inputs to the corresponding prediction models.}
    \label{fig:segments}
\end{figure}

Based on this insight, we can generalize the integration problem as follows: Given the coordinates of \( n \) points, \( (X_i, Y_i) \) for \( i = 1, 2, \ldots, n \), and the interval of integration \([X_{\text{min}}, X_{\text{max}}]\), where \textbf{\( X \) is assumed to be ordered}, we aim to solve the definite integral \( I \) over the intervals of a curve formed by this set of points, without explicitly constructing a curve:
\[
I = \int_{X_{\text{min}}}^{X_{\text{max}}} Y(X) \, \text{d}X.
\]
For BD-BR, \( X \) corresponds to \( D \) and \( Y \) to \( r \); for BD-quality, the reverse holds.

We assume that \( I \) obeys a normal distribution under the given conditions, \( I \sim P(I) = \mathcal{N}(\mu, \sigma^2) \). We employ the Maximum Likelihood Estimation (MLE) method to model the conditional probability distribution of \( I \) relative to the sample points. The likelihood function \( L(\mu, \sigma^2 \mid I) \), is defined as the joint probability of observing the integral \( I \) given the distribution parameters \( \mu \) and \( \sigma \), where \( \mu \) and \( \sigma \) are functions of \( X, Y, X_{\text{min}}, \) and \( X_{\text{max}} \). After obtaining \( P(I) \), we can determine a confidence interval for \( I \), such that \( I \) has a sufficiently high probability of falling within this interval as a quantitative measure of the confidence in the estimate. Based on the properties of the normal distribution, we choose the interval as \( [\mu - 3\sigma, \mu + 3\sigma] \). The corresponding BD values of the interval are named as Bjøntegaard Delta Confidence Interval (\(\textbf{BDCI}\)), which represents the high probability that the BD falls within the interval. The variants of this indicator on $\Delta R$ and $\Delta D$ are named ``BDCI-BR'' and ``BDCI-quality'' correspondingly. Also, the term ``quality'' is replaced with the name of the indicator when using a specific indicator, e.g. BDCI-PSNR.

We note that the number of input sample points is variable. Therefore, we segment the curve along the \( X \) values of the sample points and predict each segment separately. Specifically,
\begin{align}
    \mu = \sum \mu_i, 
    \sigma^2 = \sum \sigma_i^2,
\end{align}
where the prediction for the integral on segment \( i \) follows a normal distribution \( \mathcal{N}(\mu_i, \sigma_i^2) \).

We assume the integral over each segment obeys an independent distribution to each other and modeled using its four nearest sample points. This categorizes the segments into 7 distinct types, all of which are detailed in Figure \ref{fig:segments}. For segments that include the boundaries of the integral, an additional parameter, either \( X_{\text{min}} \) or \( X_{\text{max}} \), is required, resulting in an input size of 9. Conversely, for segments that do not include these boundaries, the input size is reduced to 8. Each type is modeled using a simple multilayer perceptron (MLP), as illustrated in Figure \ref{fig:mlp}. The network's output consists of two components, representing the predictions for \( \mu_i \) and \( \log \sigma_i \), with \( i \) indicating the index of the segment. The MLP is optimized by the log-likelihood loss:
\begin{align}
    \ell(\theta|I) = -\log f\left(I \mid \mu(\theta, \psi), \sigma^2(\theta, \psi)\right),
\end{align}
where \( \theta \) are the parameters of the MLP and \( \psi \) are the inputs to the MLP, which include the coordinates of the 4 nearest points and the \( X_{\text{min}} \) or \( X_{\text{max}} \) boundary. \( f(x \mid \mu, \sigma^2) \) is the Probability Density Function (PDF) of the normal distribution \( \mathcal{N}(\mu, \sigma^2) \).

The MLP incorporates six hidden layers, each consisting of 128 neurons, which results in an approximate total of 80,000 parameters. This network does not require significant resources from a standard CPU. It is capable of performing its calculations at a remarkably swift pace.

To improve the model's generalization, the sample point coordinates are normalized before being fed into the MLP. Specifically, the minimum and maximum values of \( X \) and \( Y \) are identified and scaled isometrically to the 0-1 range:
\begin{align}
    X' &= \frac{X - \min(X)}{\max(X) - \min(X)},  \\
    Y' &= \frac{Y - \min(Y)}{\max(Y) - \min(Y)},
\end{align}
where \( X' \) and \( Y' \) are the normalized values for \( X \) and \( Y \), respectively. This ensures that our model captures the shape information of the curve, independent of the actual range of the values.


\Section{R-D Dataset}

We constructed an extensive R-D dataset to assess the reliability of our proposed scheme. This dataset encompasses precise R-D curves for six diverse image codecs across 315 widely utilized test images. The codecs include both conventional and Variable-Bitrate (VBR) Lossy Image Codecs (LICs), as detailed in Table \ref{tab:codecs}. The images were sourced from the Kodak~\cite{kodak}, CLIC2020~\cite{clic}, and TECNICK~\cite{tecnick} datasets.

In addition, we examined seven prevalent quality assessment metrics: Root Mean Square Error (RMSE) and PSNR for signal fidelity, Structural Similarity Index 
(SSIM) \cite{wang2004ssim} and Multi-Scale Structural Similarity Index (MS-SSIM)~\cite{wang2003msssim} for structural similarity, PSNR-HVS~\cite{psnr-hvs} and PSNR-HVS-M~\cite{psnr-hvsm} for subjective human visual system metrics, and Learned Perceptual Image Patch Similarity (LPIPS)~\cite{lpips} for feature similarity.

Each image was encoded 50 times using each codec with consistent quality parameters to generate the most precise R-D curve data. This rigorous process resulted in 94,500 decoded images and 13,230 R-D curves.

Utilizing the R-D curve data from our sampling, we developed an R-D benchmark test set. Each test set sample comprises the exact anchor R-D curve paired with the target codec R-D curve. Concurrently, we conducted uniform QP-based sampling on the target codec to establish fixed sample points, with each point labeled accordingly. We sampled each benchmark R-D curve once with 4, 5, 6, 7, and 8 sample points, yielding a total of 47,688 valid sample sets. BPG served as the anchor for all samples. Our model's training process ensures that samples derived from the test set are excluded from training.

\begin{table}[t]
\centering
\caption{\textbf{Image Codecs Used in Experiments.} ``LIC (VBR)'' refers to a LIC that natively supports variable bit rates. ``LIC (retrained VBR)'' refers to a LIC to which a VBR module has been manually added and retrained. Note that such retraining incurs a performance loss.}
\small
\begin{tabular}{@{}ccc@{}}
\toprule
Name    & Type               & \# of Params \\ \midrule
JPEG~\cite{wallace1992jpeg}    & Traditional        & -           \\
BPG~\cite{bpg}     & Traditional        & -           \\
EVC~\cite{wang2023evc}     & LIC (VBR)           & 17.4M       \\
QARV~\cite{Duan2024QARV}    & LIC (VBR)           & 93.4M       \\
LIC-TCM~\cite{liu2023tcm} & LIC (retrained VBR) & 160.5M      \\
MLIC++~\cite{jiang2023mlic++}  & LIC (retrained VBR) & 116.7M      \\ \bottomrule
\end{tabular}
\label{tab:codecs}
\end{table}

\Section{Experimental Results}

\begin{figure}[t]
    \centering
     \begin{subfigure}[b]{0.48\textwidth}
         \centering
         \includegraphics[width=\textwidth]{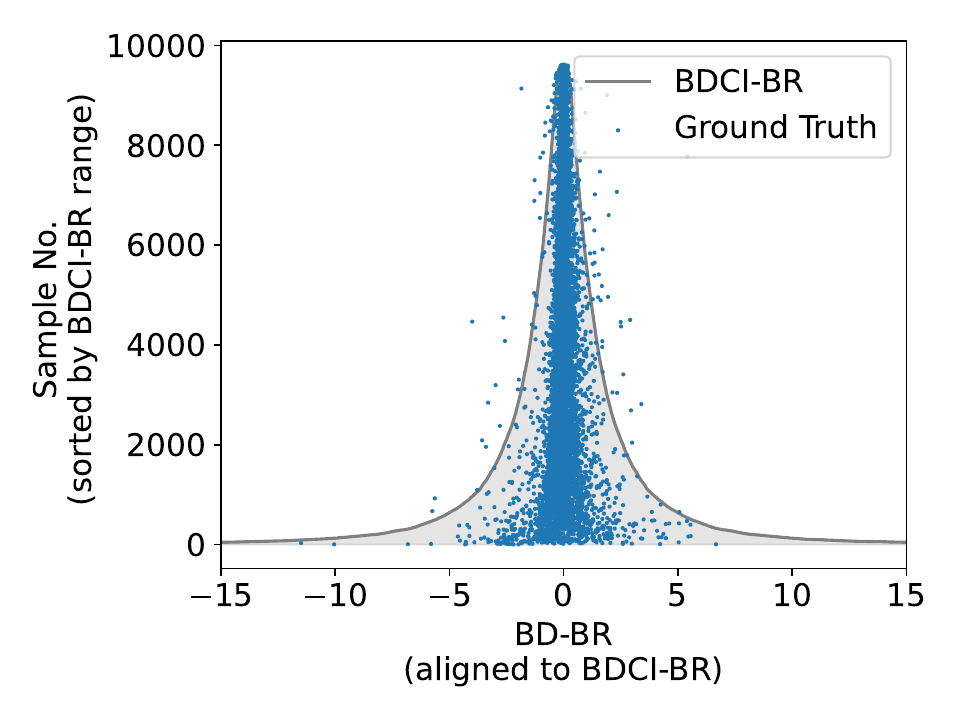}
         \caption{}
     \end{subfigure}
     \begin{subfigure}[b]{0.48\textwidth}
         \centering
         \includegraphics[width=\textwidth]{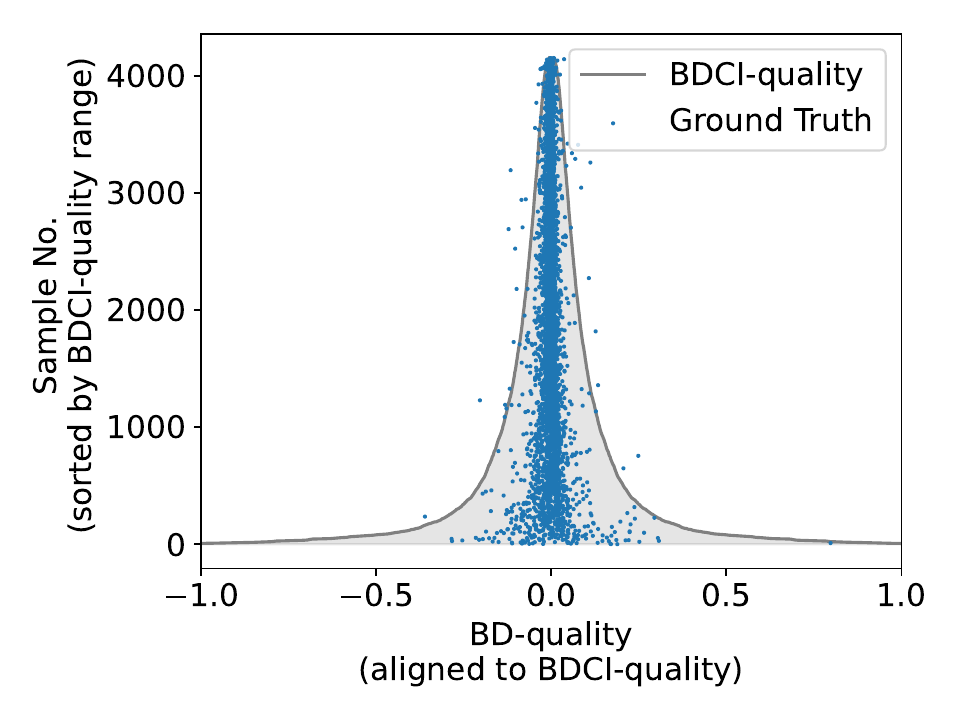}
         \caption{}
     \end{subfigure}
    \caption{\textbf{BDCI vs. Ground Truth.} For visualization purposes, we aligned the centroids of the BDCI and sorted them by the length of the BDCI intervals. The right plot only shows the results for the PSNR, PSNR-HVS, and PSNR-HVS-M metrics because the scales of the other metrics could not be aligned.}
    \label{fig:BDCI-vs-gt}
\end{figure}
\begin{figure}[h]
    \centering
     \begin{subfigure}[b]{0.45\textwidth}
         \centering
         \includegraphics[width=\textwidth]{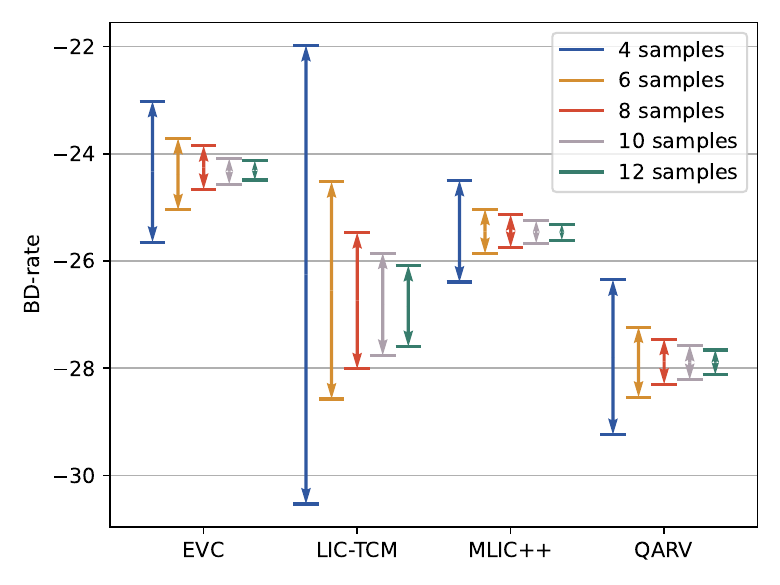}
         \caption{BD-BR}
     \end{subfigure}
     \begin{subfigure}[b]{0.45\textwidth}
         \centering
         \includegraphics[width=\textwidth]{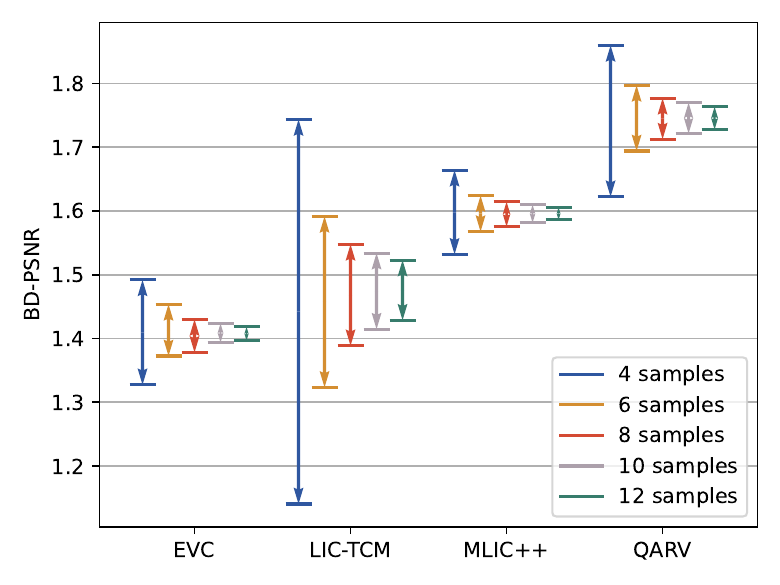}
         \caption{BD-quality}
     \end{subfigure}
    \caption{\textbf{BDCI vs. R-D Sample Size.} The anchor codec is BPG. As the number of R-D sample points increases, the BDCI intervals become narrower.}
    \label{fig:468points}
\end{figure}

We conducted an evaluation to ascertain the efficacy of the BDCI metric within our dataset. Figure \ref{fig:BDCI-vs-gt} illustrates the correlation between BDCI and the actual BD values when utilizing four sample points for the test set, with the data normalized for clarity. The results indicate that BDCI accurately predicts the BD confidence intervals in the majority of instances, with less than 1\% of ground truth BD values lying outside the BDCI.

To verify that the BDCI decreases as the number of sample points increases, we tested the results using the BDCI metric on a variety of encoders. Figure \ref{fig:468points} depicts the effect of varying the number of sample points on BDCI accuracy. In the default case of 4 sample points, the width of the BDCI is relatively large, indicating that the BD measurement is less reliable at this point. As the number of sample points increases, the width of the BDCI gradually narrows and eventually converges. This is a good indication that the BDCI provides a reasonable response as the R-D samples tend to become more accurate.

\begin{figure}[h]
    \centering
    \begin{subfigure}[b]{0.31\textwidth}
        \includegraphics[width=\linewidth]{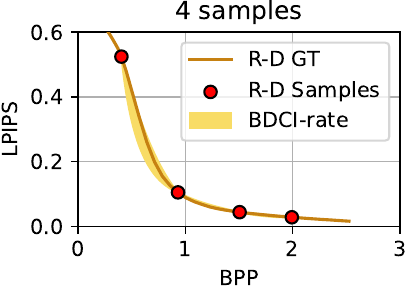}
    \end{subfigure}
    \begin{subfigure}[b]{0.31\textwidth}
        \includegraphics[width=\linewidth]{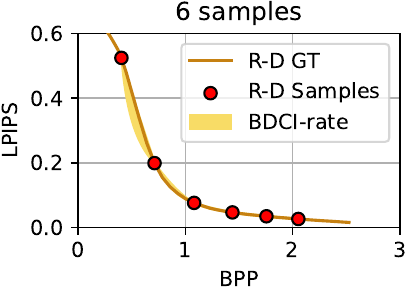}
    \end{subfigure}
    \begin{subfigure}[b]{0.31\textwidth}
        \includegraphics[width=\linewidth]{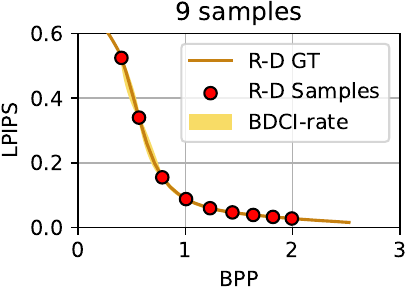}
    \end{subfigure}
    \caption{\textbf{BDCI Visualize.} The BD-BR of the upper and lower bounds of the yellow area are equal to the upper and lower bounds of the corresponding BDCI-BR respectively. ``R-D GT'' indicates the ground truth R-D curve. We hide the Anchor R-D curve for better visualization.}
    \label{fig:bdci_visualize}
\end{figure}

We visualize the results of the BDCI calculations in Figure \ref{fig:bdci_visualize}, using the same R-D samples as in Figure \ref{fig:sample-lpips}. The BD-BR of the upper and lower boundaries of the yellow region in the figure is equal to the upper and lower boundaries of the BDCI. Please note that this region does not actually exist during the computation and is drawn manually for the purpose of visualizing the BDCI. The results clearly demonstrate that BDCI's calculations are reasonable and consistent with human intuition.

Furthermore, we employed the $\mu$ output from our model as a direct BD estimator and compared its performance against prevalent BD estimation methods using $4\sim8$ R-D sample points. The experimental outcomes are presented in Table \ref{table:estimation_error}, where the estimation bias is quantified using the Mean Square Error (MSE), representing the discrepancy between the actual BD values and the BD predictions based on the sample points. The experimental findings conclusively show that our proposed approach achieves superior prediction accuracy in most of the conditions, especially when calculating BD-BR.

\begin{table}[t]
\centering
\caption{\textbf{Comparison of Estimation Bias:} Mean Squared Error (MSE) for BD estimation using 4 to 8 sample points with our proposed method compared to commonly used BD estimation methods. The errors in BD-BR and BD-quality are indicated by \(\epsilon(\Delta R)\) and \(\epsilon(\Delta D)\), respectively. \(\epsilon(\Delta D)\) is scaled for better presentation.}
\small
\begin{tabular}{@{}cc|ccccccc@{}}
\toprule
\multicolumn{2}{c|}{}                         & PSNR           & MS-SSIM        & LPIPS          & RMSE           & SSIM           & \begin{tabular}[c]{@{}c@{}}PSNR-\\ HVS\end{tabular} & \begin{tabular}[c]{@{}c@{}}PSNR-\\ HVSM\end{tabular} \\
\multicolumn{2}{c|}{\begin{tabular}[c]{@{}c@{}}Order of \\ magnitude\end{tabular}} & \begin{tabular}[c]{@{}c@{}}$\times 1$\\ $\times 10^{-3}$\end{tabular}         & \begin{tabular}[c]{@{}c@{}}$\times 1$\\ $\times 10^{-7}$\end{tabular}         & \begin{tabular}[c]{@{}c@{}}$\times 1$\\ $\times 10^{-7}$\end{tabular}         & \begin{tabular}[c]{@{}c@{}}$\times 1$\\ $\times 10^{-9}$\end{tabular}         & \begin{tabular}[c]{@{}c@{}}$\times 1$\\ $\times 10^{-7}$\end{tabular}         & \begin{tabular}[c]{@{}c@{}}$\times 1$\\ $\times 10^{-3}$\end{tabular}                                              & \begin{tabular}[c]{@{}c@{}}$\times 1$\\ $\times 10^{-3}$\end{tabular}                                               \\ \midrule
\multirow{2}{*}{CSI}                         & $\epsilon(\Delta R)$      & 0.560          & 1.027          & 199.884        & 0.772          & 73.641         & 0.627                                               & 0.647                                                \\
                            & $\epsilon(\Delta D)$      & 0.500          & 0.060          & 10.451         & 1.052          & 0.322          & 0.956                                               & 2.326                                                \\ \midrule
\multirow{2}{*}{PCHIP}      & $\epsilon(\Delta R)$      & 0.106          & 0.129          & 0.686          & 0.101          & 0.155          & 0.120                                               & 0.191                                                \\
                            & $\epsilon(\Delta D)$      & 0.245          & \textbf{0.032} & \textbf{6.671} & \textbf{0.365} & 0.284          & 0.372                                               & \textbf{0.902}                                       \\ \midrule
\multirow{2}{*}{Akima}      & $\epsilon(\Delta R)$      & -              & 0.160          & 3.870          & 0.167          & 0.496          & 0.169                                               & 0.229                                                \\
                            & $\epsilon(\Delta D)$      & 0.266          & 0.040          & 6.626          & 0.379          & 0.290          & 0.401                                               & 0.982                                                \\ \midrule
\multirow{2}{*}{Ours}       & $\epsilon(\Delta R)$      & \textbf{0.104} & \textbf{0.115} & \textbf{0.566} & \textbf{0.083} & \textbf{0.154} & \textbf{0.120}                                      & \textbf{0.191}                                       \\
                            & $\epsilon(\Delta D)$      & \textbf{0.224} & 0.054          & 6.772          & 0.413          & \textbf{0.273} & \textbf{0.353}                                      & 1.022                                                \\ \bottomrule
\end{tabular}
\label{table:estimation_error}
\end{table}

We measured the running time of the algorithm, as presented in Table \ref{table:runtime}. The findings from experiments conducted with varying numbers of R-D sample points reveal that, despite integrating neural networks into our approach, the runtime for a single execution remains minimal on a single-core CPU.

\begin{table}[h]
\centering
\caption{\textbf{Execution Time.} Evaluated on a single core of an Intel$^{\text{\tiny\textregistered}}$ Xeon$^{\text{\tiny\textregistered}}$ Silver 4310 CPU operating at 2.10GHz.}
\begin{tabular}{@{}cccc@{}}
\toprule
Number of R-D Samples & 4   & 6    & 8    \\ \midrule
Runtime [ms]          & 6.1 & 11.6 & 14.3 \\ \bottomrule
\end{tabular}
\label{table:runtime}
\end{table}

\FloatBarrier

\Section{Suggestions \& Conclusion}

In this study, we identified a critical flaw in the existing BD estimation methodology and offer a novel deep learning-based solution. Our method not only achieves superior accuracy in estimating the posterior probability distribution of the BD integral but also provides reliable confidence interval estimates. The experimental findings suggest avenues for future research aimed at further minimizing estimation errors and enhancing the robustness of BD estimation outcomes.

Our study suggests that the BD comparison method may be flawed when the R-D sample size is limited, which could significantly affect the reliability of the results. To enhance the accuracy and reliability of BD comparisons, we recommend the following:

\begin{enumerate}
    \item \textbf{Implement BDCI Metric.} Use the methodology from this paper for codec comparisons, including BD estimates and BDCI intervals.
    \item \textbf{Increase Sample Points.} Use more sample points for BD metrics to enhance accuracy and reduce BDCI intervals.
    \item \textbf{Use Dense Sample Points with Anchor Method.} When using fast codecs as anchor, obtain a dense R-D anchor curve for better measurement accuracy.
\end{enumerate}

\FloatBarrier
\Section{References}
\bibliographystyle{IEEEbib}
\bibliography{refs}

\end{document}